\documentstyle[wds11,epsf]{article}

\lefthead{BESHLEY, PETRUK}
\righthead{HADRONIC IMAGES OF SUPERNOVA REMNANTS}

\begin{document}

\title{The properties of gamma-ray images of supernova remnants due to proton-proton interactions}

\author{V. Beshley, O. Petruk}

\affil{Institute for Applied Problems in Mechanics and Mathematics, Lviv, Ukraine}

\begin{abstract}
MAGIC and H.E.S.S experiments are the first to produce images of supernova remnats (SNRs) in TeV gamma-rays. The gamma-radiation are produced either by electrons (due to inverse-Compton scatterings) or protons (due to pion decays). We present a method to synthesize gamma-ray images of Sedov SNRs due to hadronic emission. The model is developed in the frame of a classic approach to proton acceleration and hydrodynamics of the shocks in a uniform interstellar medium; it includes energy losses of relativistic protons due to pp interactions. Our calculations show that these losses are important only for large densities of protons as it could be in case of interactions of the supernova shock with molecular cloud. Numerical simulations are used to synthesize radial profiles of hadronic TeV gamma-rays.
\end{abstract}

\begin{article}

\section{Introduction}
Cosmic rays (CRs) are the widely studied. Supernova remnants (SNRs), main source of galactic CRs, are excellent objects to study magneto-hydrodynamics of nonrelativistic shocks and acceleration of cosmic rays, namely protons and electrons. These particles radiate from radio to $\gamma$-rays due to different types of emission. Experiments in high-energy astronomy observe all types of these emission.

Most of galactic cosmic rays are belived to be produced by the forward shocks in SNRs. In particular, efficient proton acceleration changes the structure of the shock front and makes plasma more compressible that leads to lower adiabatic index, to increased shock compression factor and to some observed effects: reduced physical separation between the forward shock and the ``contact discontinuity'' (or reverse shock) [\markcite{{e.g \it Warren et al.}, 2006}]; concave shape of the energy spectrum [\markcite{{e.g \it Reynolds and Ellison}, 1992}]; growth of some turbulence modes and to MF amplification in the pre-shock region [\markcite{{e.g \it Bell}, 2004}]; ``blinking'' X-ray spots originated from such growth of MF [\markcite{{\it Uchiyama et al.}, 2007}]. 

Observations are expected to confirm that protons are accelerated in SNRs to very high energy and emit (TeV) $\gamma$-rays. Nevertheless, analysis of the broad-band spectra of SNRs shows that both electrons and protons may be responsible for TeV gamma-rays [\markcite{{e.g. SN 1006:\it Acero et al.}, 2010}].

The properties of the thickness of the radial profiles of hard X-ray brightness are used to estimate the strength of the post-shock magnetic field [\markcite{{\it Berezhko et al.}, 2003}]. Radial profiles of the radio brightness may constrain the time evolution of the electron injection efficiency [\markcite{{\it Petruk et al.}, 2011a}]. In a simple fashion, the radial profiles of hadronic TeV $\gamma$-ray brightness are sensitive to the density of ambient medium. This property is the subject of the present study.

Properties of the nonthermal images of Sedov SNRs due to radiation of accelerated electrons in radio, X-rays and $\gamma$-rays are systematically studied in [\markcite{{\it Reynolds}, 1998, 2004}] and [\markcite{{\it Petruk et al.}, 2009, 2011b, Papers I and II respectively}]. Numerical models for synthesis of maps of adiabatic SNRs in uniform ISM and and uniform interstellar magnetic field (ISMF) from basic theoretical principles as well as their approximate analytical descriptions are developed in these papers. The main factors determining the azimuthal and radial variation of surface brightness of Sedov SNRs are determined there. 

These papers, as the present one, are limited to the test-particle approach because the non-linear theory of diffusive acceleration is not developed for shocks of different obliquity, while the obliquity dependence of various parameters is important for image modelling. 

In the present paper, we study properties of the radial profiles of surface brightness in $\gamma$-ray due to proton-proton interactions including of the energy losses of proton.

\section{Model}
Our model closely restores that used in Papers I and II. 
Let us consider an adiabatic SNR in uniform interstellar medium (ISM) and uniform interstellar magnetic field (ISMF). We use quite accurate approximate formula in Lagrange coordinates [\markcite{{\it Petruk}, 2000}] for description of hydrodynamics of SNRs in the adiabatic stage of evolution (Sedov solutions [\markcite{{\it Sedov}, 1959}]). Magnetic field is described following [\markcite{{\it Reynolds}, 1998}]. We do not consider amplification of the ambient field.

At the shock, the spectrum of acceleration protons is taken as 
\begin{equation}
N_{p}(E_{p}) =K_{s}E_{p}^{-s}\exp\left(-\frac{E_{p}}{E_{p,max}}\right)
\label{spectr}
\end{equation}
where $E_{p,max}$ is the maximum energy of protons, $s$ is constant and we use $s=2$.

We assume that variation of the maximum energy with obliquity angle (angle between the ambient magnetic field and the shock velocity) is constant and the injection efficiency of protons is isotropic. 

The surface brightness is calculated integrating emissivities along the line 
of sight within SNR. 
Hadronic $\gamma$-rays appear as a consequence of the neutral pion and $\eta$-meson decays produced in inelastic collisions of 
accelerated protons with {\em thermal protons downstream of the shock}; the spatial distribution of the target protons is simply 
proportional to the plasma density. 
The hadronic $\gamma$-ray emissivity is calculated as [\markcite{{\it Kelner et al.}, 2006}]
\begin{equation}
 q_{\gamma}(E_{\gamma})=cn_{H}\int\limits_{0}^{1} \sigma_{pp}(E_{\gamma}/x) N(E_{\gamma}/x)
 F_\gamma(x,E_{\gamma}/x)\frac{dx}{x},
 \label{pp-emiss}
\end{equation}
where $x=E_{\gamma}/E_{p}$, the cross-section is [\markcite{{\it Aharonian, Athoyan}, 2000}]:
\begin{equation}
\sigma_{pp}(E_{p})=28.5+1.8\ln{(E_{p}/1\ {GeV})} \ \  \ {mb},
\label{across}
\end{equation}
the function $F_{\gamma}$ is [\markcite{{\it Kelner et al.}, 2006}]
\begin{equation}
\begin{array}{l}
 \displaystyle
F_{\gamma}(x,E_{p})
=B_{\gamma}\frac{\ln{(x)}}{x}\left(\frac{1-x^{\beta_{\gamma}}}{1+k_{\gamma}x^{\beta_{\gamma}}(1-x^{\beta_{\gamma}})}\right)^{4}
\\ \\ \displaystyle\qquad
\times\left[\frac{1}{\ln{(x)}}-\frac{4\beta_{\gamma}x^{\beta_{\gamma}}}{1-x^{\beta_{\gamma}}}-\frac{4k_{\gamma}\beta_{\gamma}x^{\beta_{\gamma}}(1-2x^{\beta_{\gamma}})}{1+k_{\gamma}x^{\beta_{\gamma}}(1-x^{\beta_{\gamma}})}\right],
\end{array}
 \label{eq58}
\end{equation}
where
\begin{equation}
B_{\gamma}=1.30+0.14L+0.011L^{2},
 \label{eq59}
\end{equation}
\begin{equation}
\beta_{\gamma}=\frac{1}{1.79+0.11L+0.008L^{2}},
 \label{eq60}
\end{equation}
\begin{equation}
k_{\gamma}=\frac{1}{0.801+0.049L+0.014L^{2}}.
 \label{eq61}
\end{equation}
and $L=\ln(E_{p}/1\ \ {TeV})$.
\subsection{Energy losses of protons due to proton-proton interactions}

Modelling the surface brightness distribution and maps due to proton collisions is important to accurate the energy losses due to pion productions. The losses due to proton collisions are important for higher densities of target protons. It may be shown the proton collision losses may be described as
\begin{equation}
-\left(\frac{dE_{p}}{dt}\right)_{pp}=3\kappa cn_{H}\sigma_{pp}(E_{p})E_{p,kin},
 \label{Bloss3_3}
\end{equation}
where the factor $3$ accounts for the production of $\pi^{0}$, $\pi^{+}$ and $\pi^{-}$ mesons, respectively and $c$, $n_{H}$ are speed of light, the proton target density, $\sigma_{pp}(E_{\pi},E_{p})$  the differential cross-section for the interaction of two protons, $E_{p}$ the energy of primary proton, $3\kappa=0.51$ if $\kappa=0.17$ [\markcite{{\it Aharonian et al.}, 2000}]. 

Let us compare the energy losses due to proton-proton interactions with the radiative losses of electrons. The losses of electrons are given by
\begin{equation}
-\left(\frac{dE_{e}}{dt}\right)_{rad}=\frac{4}{3}\sigma_{T}c\left(\frac{E}{m_{e}c^{2}}\right)^{2}\left(\frac{B^{2}}{8\pi}\right).
 \label{ploss}
\end{equation}
where $\sigma_{T}$ is the Thomson cross-section, $m_{e}$ the mass of electron. 
The ratio of electron to proton radiative losses is
\begin{equation}
\frac{\dot{E}_{e,rad}}{\dot{E}_{p,pp}}\simeq 5\frac{B^{2}_{\mu G}E^{2}_{e,TeV}}{n_{H}E_{p,TeV}},
\label{ploss}
\end{equation}
where we used $\sigma_{pp}\approx 33\ {mb}$, $B_{\mu G}$ is the magnetic field in $10^{-6} G$, and $E_{e,TeV}$ and $E_{p,TeV}$ are the energy of electrons and protons in $10^{12} eV$. The losses of electrons end protons is similar when the density of protons $440\ cm^{-3}$ and the typical galactic magnetic field $B_{\mu G}=3$ the maximum energy of electrons end protons are $30$ and $100$ TeV respectively. 
One can see that the losses of protons with energy $1000\ TeV$ are comparable to losses of electrons with energy $30$ TeV in magnetic field $3\ {\mu G}$, if the number density of target protons is rather high $440\ {cm^{-3}}$. 

\subsection{Downstream evolution of the proton distribution}

Let the energy of proton at the time $t_{i}$, when it leave the region of acceleration, is $E_{pi}$. Then it is smaller at the present time $t$,
\begin{equation}
 E_{p}={E_{pi}}{{\cal E}_{ad}(\bar{a})^{\mu(\bar{a})}{\cal E}_{pp}(E_{p},\bar{a})}, 
\end{equation}
because the terms responsible for the adiabatic ${\cal E}_{ad}$ and collisional ${\cal E}_{pp}$ losses are equal or smaller than unity; $\bar a=a/R$, $a$ the Lagrangian coordinate, $R$ the radius of SNR,
\begin{equation}
 {\cal E}_{ad}(\bar{a})=\bar n(\bar a)^{1/3},
 \label{ppmaps:adloss}
\end{equation}
where 
$\bar n=n/n_{s}$, 
index ``s'' denotes the value immediately post-shock, 
\begin{equation}
 {\cal E}_{pp}(E_{p},\bar{a})=\left(E_{p}/1{GeV}\right)^{1-\mu(\bar{a})}{\cal I}(\bar{a}),
 \label{adpploss_t}
\end{equation}
$\mu(\bar{a})$ and ${\cal I}(\bar{a})$ are dimensionless self-similar functions 
\begin{equation}
 \mu(\bar{a})=\exp\left[\zeta{\int\limits_{\bar{a}}^{1}x^{3/2}p\left( \frac{\bar{a}}{x}\right) dx}\right],
 \label{mu}
\end{equation}
\begin{equation}
I(\bar{a})=\exp\left[ \zeta\int\limits_{\bar{a}}^{1}x^{3/2}q\left( \frac{\bar{a}}{x}\right)\mu\left( \frac{\bar{a}}{x}\right) dx\right].
\label{i}
\end{equation}
where $\zeta=5tc_{1}/2=1.21\cdot 10^{-6}t_{3}n_{Hs}$, $t_3=t/1000{yrs}$. It is clear from here that ${\cal E}_{pp}$ is effective only where the density of target protons is large, at least $n_{Hs}\sim 10^6{cm^{-3}}$.

The energy spectrum of protons downstream of the shock evolves self-similarly
\begin{equation}
\begin{array}{l}
 \displaystyle
N_{p}(E_{p},\bar a,t)=K(\bar a,t)E^{-s}_{p}\ \mu(\bar a){\cal E}_{pp}(E_{p},\bar{a})^{s-1}
\\ \\ \displaystyle\qquad
\times\exp\left[ -\left(\frac{E_{p}\bar{a}^{3q/2}}{E_{p,max}{\cal E}_{ad}(\bar{a})^{\mu(\bar{a})}{\cal E}_{pp}(E_{p},\bar{a})} \right)^{\alpha}\right] .
\end{array} 
 \label{specev_}
\end{equation}
with $K(\bar a,t)=K_{s}\bar{K}(\bar{a})$, $\bar{K}(\bar{a})=\bar{a}^{3b/2}\bar{n}(\bar{a})^{1+\mu(\bar{a})(s-1)/3}$.
The downstream distribution of relativistic protons are modified by the adiabatic expansion of SNR. 
Losses due to inelastic collisions affects the distribution only when $\zeta$ is not small, i.e. when $n_{Hs}$ is large. 

\section{Results}
Analysis of the formula (\ref{Bloss3_3}) shows that the thickness of radial profiles of the TeV $\gamma$-ray surface brightness distributions are only function of density of target protons. If density of target protons is small then we can neglect the energy losses due to proton-proton interactions. However, when the density of target protons is more than $\sim 400\ cm^{3}$ one can not neglect these losses. However,they reveal themselves in the radial profiles for even higher densities.
\begin{figure}[htb]
\begin{center}
\begin{tabular}{c}
  \epsfxsize=82mm
  \epsfbox{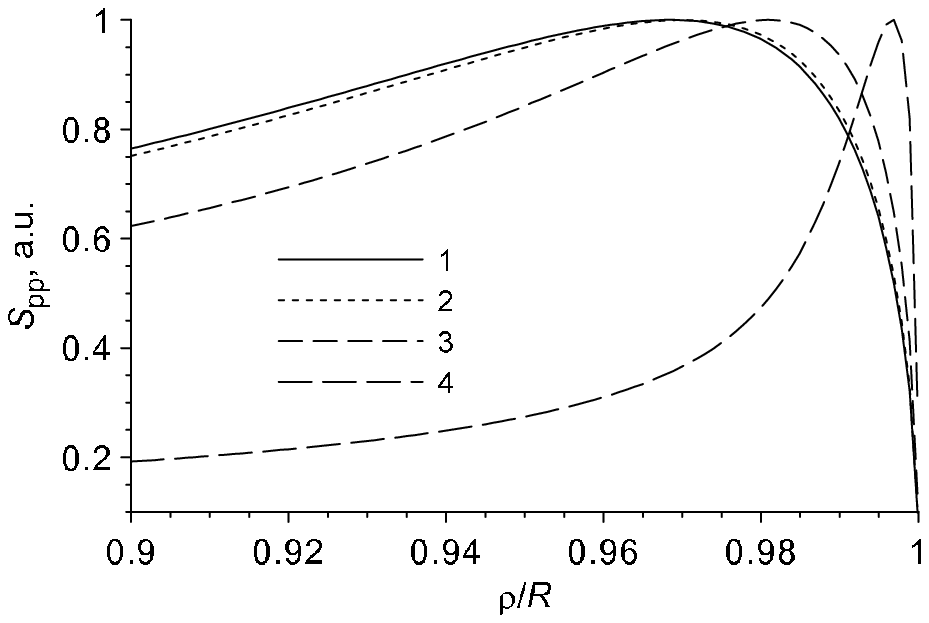}
\end{tabular}
\end{center}
\caption{Radial profiles of the $\gamma$ -ray surface brightness due to hadronic emission for different densities of the target protons: $1cm^{-3}$ (line 1), $10^{4}cm^{-3}$ (line 2), $10^{5}cm^{-3}$ (line 3), $10^{6}cm^{-3}$ (line 4). $E_{max}=1000$ TeV, energy of gamma-rays is $1$ TeV.}
\end{figure}

In the fig.1, we show the influence of target protons density on the thickness of radial profiles of surface brightness. If the density increases the energy losses of protons increase as well and the thickness of the radial profile decreases. Since the downstream density is proportional to the pre-shock density, thus may be used as a probe of the ambient medium density.

\section{Conclusion}

We consider an influence of target protons density on the radial profiles of TeV $\gamma$-ray surface brightness. Downstream evolution of the proton distribution taking into account the energy losses due to proton-proton interactions is described. We show the influence of the target protons density on properties of radial profiles of $\gamma$-ray brightness distribution. These properties can be used to estimate the density of molecular clouds that interact with the shock wave of SNR.

\end{article}

\end{document}